\documentclass[12pt]{iopart}
\usepackage[english]{babel}
\usepackage{graphicx}
\usepackage{xcolor}
\begin{document}

\title[Cooling of an objetct by...]{Cooling of an objetct by forced convection}

\author{Mateo Dutra$^{1a}$, Martín Monteiro$^{2b}$, Arturo C. Martí $^{1c}$}

\address{$^1$Instituto de Física, Universidad de la República, Montevideo, Uruguay; $^2$Universidad ORT Uruguay, Montevideo, Uruguay}
\ead{$^a$mateodutrafisica@gmail.com, $^b$fisica.martin@gmail.com,$^c$marti@fisica.edu.uy}
\vspace{10pt}
\date{\today}

\begin{abstract}
We present an experiment on forced convection where a previously heated object is cooled  under the effect of a controlled stream of air.  We consider a square copper plate in which temperature variations can be considered negligible and we measure the cooling rate as a function of the average velocity of the air stream. We use a thermal camera to measure the temperature field 
and the  cooling curves as a function of time for different conditions. An empirical relation between the characteristic cooling time and the mean velocity of the air stream is reported. The results obtained are discussed in the framework of simple dimensional models and their limits of validity.
\end{abstract}

\section{Introduction}

In thermodynamics, temperature variations in a system related to energy flows within the system and with the environment are of fundamental importance. In general, three heat transfer mechanisms are distinguished: conduction, radiation and convection \cite{bergman2011fundamentals}. Energy transfers by conduction typically occur within a solid due to temperature differences within it and can be quantified by means of the thermal conductivity, the area of the object and the temperature difference. The second mechanism mentioned, thermal radiation, is due to the conversion of thermal energy of the material charges into electromagnetic radiation. It  depends on the temperature of the object, its emissivity, its area and the ambient temperature. Finally, convection, also known as convective heat transfer, is generally defined as the transfer of heat between two sites due to the movement of the fluid. This heat transfer mechanism is usually predominant in liquids and gases and
in many experimental situations its effect overcomes that of the other heat transfer mechanisms. The importance of these phenomena covers all scales from the smallest to the astrophysical, including geophysical and climatic phenomena.

The cooling of an object placed in front of an air stream is in principle a complex phenomenon that depends on many factors.  A model that claims to be accurate should take into account the thermodynamic properties and variables of both the object and the jet fluid as well as the relative orientation. In this paper we  focus on a situation where we reduce the variables to be considered to a minimum. It is worth mentioning that a similar problem was proposed by Lord Rayleigh in the early days of dimensional analysis in a paper \cite{rayleigh1915principle} where he proposed several problems as an application of the then recent Buckingham's Pi theorem. The proposed solution was not without controversy, as can be found at \cite{bergman2011fundamentals}.

Within the great variety of phenomena related to heat transfer, situations where an object is cooled occupy a prominent place. Newton's law of cooling states that if the temperature differences between the object and the environment are not very large, a law of exponential type is verified \cite{vollmer2009newton}. In instructional laboratories for undergraduate students several  experiments have been proposed  centered on this law and the cooling of objects in different conditions \cite{sullivan1990newton,spuller1993cooling,planinvsivc2008surface}. Others experiments have focused on natural convection phenomena where the motion is due to buoyancy forces caused by variations of the fluid density with temperature. In contrast, situations where convection is forced, i.e., the fluid is forced by an external mechanism and impinges on the cooling object, have received very little attention \cite{gonzalez2008investigating}.

In this work we study the cooling of an object in the presence of an externally controlled air stream with the objective of determining the exponent that characterizes the cooling curve as a function of the average air velocity. For this purpose, we set up a square copper plate, whose temperature can be considered uniform and the heat radiation mechanisms are negligible, and we studied the cooling in the presence of an incident air stream. The temperature of the plate is obtained by using a thermal camera attached to a smartphone. This tool has proven to be 
useful in several  demonstrative  or instructional laboratory experiments \cite{haglund2015thermal,mollmann2007infrared,guido2023seeing}. Since the price has dropped in the last years \cite{vollmer2018infrared}, they are being more frequently used and different activities have been proposed in many physics topics \cite{kacovsky2019electric,oss2015electro,xie2011infrared,cabello2006infrared,gfroerer2015thermal,Pendrill_2024}. 
The development of IR cameras that work in conjunction with smartphones opens up the possibility of new experiments at lower cost.
\cite{monteiro2022resource}. Other innovative alternatives recently proposed in the literature for the experimental study of cooling laws are based on the use of Arduino microcontrollers \cite{Galeriu2018arduino}.
Below we briefly summarize the theoretical framework of this experiment. In section III we discuss the experimental setup. The results are presented and discussed in section IV and finally in section V we formulate the closing remarks.

\section{Theoretical framework}

In the realm of heat transfer, various mechanisms operate concurrently, each exerting its influence on the thermal dynamics of a system. Yet, there are instances where one mode of heat transfer predominates over others, shaping the overall heat exchange process. To quantify the significance of each heat transfer mechanism, it is usual to employ non-dimensional numbers, providing a framework to assess their relative impacts within a given context.

The dimensionless Biot number, $\mathrm{Bi}$, is defined by the ratio between the conductive and the convective heat transfer coefficients. When the Biot number verifies $\mathrm{Bi} \ll 1$ the internal flow is much larger than the heat loss from the surface, so we can consider the object as homogeneously heated \cite{vollmer2009newton}. In this case, and if the radiative cooling contribution is neglected, the temperature $T_o$ of the object located in a medium with temperature $T_e$ varies over time $t$ according to
\begin{equation}
    mc\frac{dT_{o}}{dt}=-\alpha_{conv}\cdot A\cdot (T_{o} - T_{e})
    \label{eq1}
\end{equation}
where $m$ is the mass of the object, $c$ is the specific heat, $\alpha_{conv}$ the convective heat transfer coefficient and $A$ the area of the object \cite{vollmer2009newton}. A solution for this equation is
\begin{equation}
    T_{o}(t) = T_{e} + (T_{i} - T_{e})\cdot e^{-\frac{t}{\tau}}
    \label{eq2}
\end{equation}
where $T_i$ is the temperature of the object when $t=0$, and $\tau$ is the characteristic cooling time defined by
\begin{equation}
    \tau = \frac{mc}{A\alpha_{conv}}.
    \label{eqTau}
\end{equation}
obtained from the temporal evolution of the temperature linearized as
\begin{equation}
    \log(T_{o}-T_{e}) = \log(T_i-T_{e}) - \frac{t}{\tau}.
    \label{eqlin}
\end{equation}
It is worth mentioning that 
in a situation where the radiation contribution is significant we must  add a term to Eq.~\ref{eq1} to become
\begin{equation}
    mc\frac{dT_{o}}{dt}=-\alpha_{conv}\cdot A\cdot (T_{o} - T_{e}) - \varepsilon\cdot\sigma\cdot A\cdot (T_{o}^4-T_{e}^4)
    \label{eqRadiation}
\end{equation}
where $\varepsilon$ is the object emissivity, and $\sigma$ the Stefan-Boltzmann constant.

Here, we are undertaking a study on the cooling process of an object exposed to an air current. The cooling behavior heavily relies on the geometry of the object, making it challenging to model accurately in the absence of symmetry. As a result, we have opted to focus on analyzing the cooling of a vertically oriented square plate positioned within a horizontal air stream that runs perpendicular to the plate's surface. For this study, 
we consider that the cross-sectional area occupied by the air stream covers the entire plate and that the air velocity can be considered constant in the region close to the plate.
In this case the most relevant geometrical characteristic of the plate is the thickness while the other dimensions do not play an important role in a first approximation.

By considering additional dimensionless numbers, specifically the Prandtl and Reynolds numbers, we can establish a correlation between the convective heat transfer coefficient and the mean  velocity of the stream $v$. The Prandtl number $\mathrm{Pr}$ is defined as the ratio of momentum diffusivity to thermal diffusivity, and the Reynolds number $\mathrm{Re}$ as the ratio of inertial forces to viscous forces. For laminar flow, if
\begin{equation}
\mathrm{Pr} > 0.6,
\label{eqpr}
\end{equation}
then $\alpha_{conv}\propto \mathrm{Re}^{1/2}$, and consequently $\alpha_{conv}\propto v^{1/2}$ \cite{bergman2011fundamentals}. In addition, if we neglect the radiation heat transfer, we obtain from Eq.~\ref{eqTau} 
\begin{equation}
    \tau \propto v^{-1/2}
    \label{eqVair}
\end{equation}
 establishing a relationship between the characteristic cooling 
 time and the air stream velocity.

In this work we use a thermal camera to measure the temperature of the vertical plate
as a  function of time of a vertical plate under the effect of forced convection originated
by an air stream. For each  air velocity we calculate  the characteristic cooling time 
$\tau$  and compare with the previous scaling relation given by Eq.~\ref{eqVair}.

\section{Experimental configuration}
\begin{figure}[htbp]
    \centering
    \includegraphics[width=.85\columnwidth]{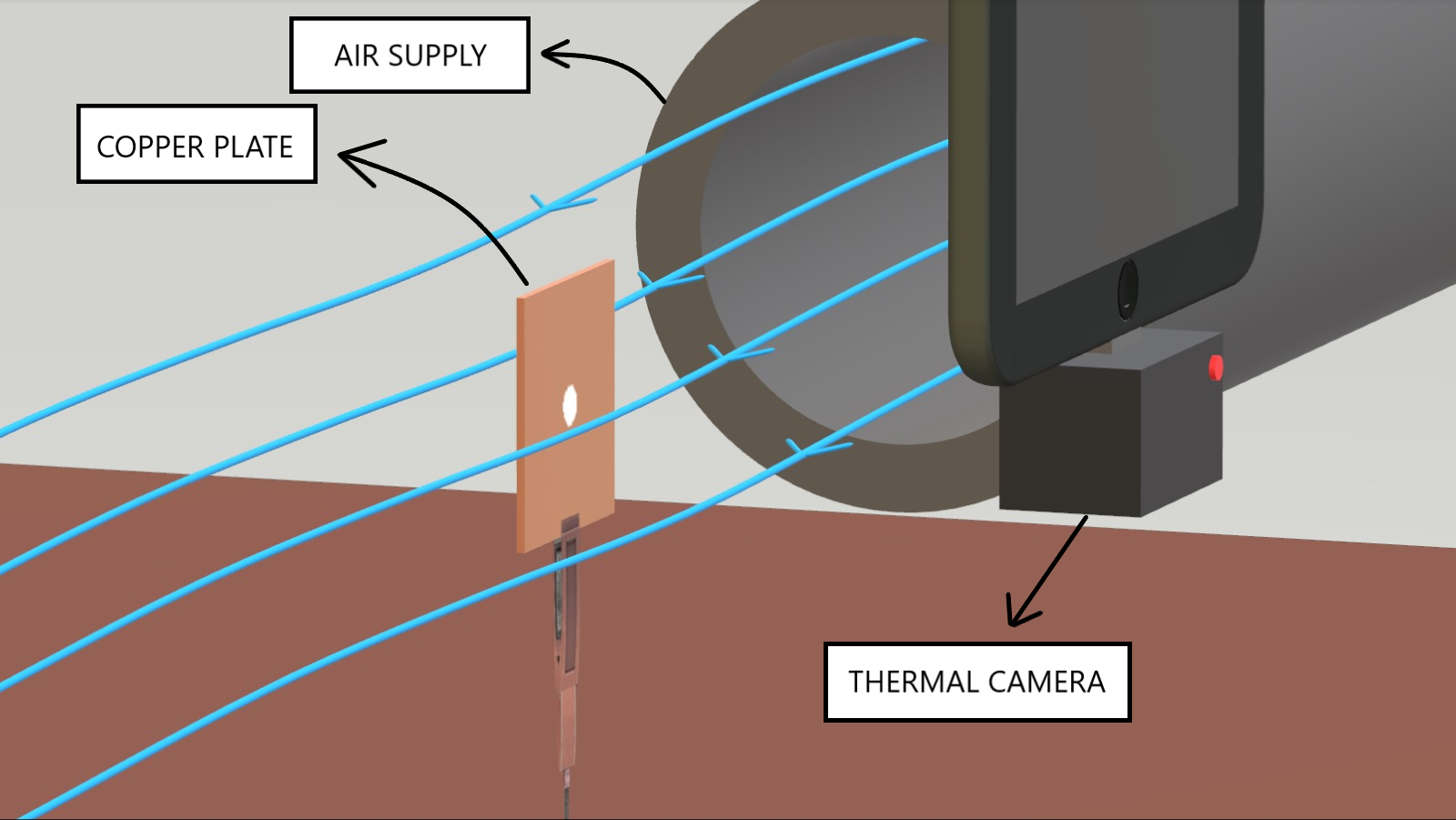}
    \caption{Experimental configuration. A plate is cooled by a horizontal airflow, and a thermal camera measures its temperature. A white circle in the center of the plate was painted 
    with white-out correction fluid, to ensure accurate measurements.
    }
    \label{fig:montaje}
\end{figure}

The experimental setup shown in Fig.~\ref{fig:montaje}  consists of a square copper plate with size 
$2.0(1) \times 2.0(1) \times 0.092(2)$ cm$^3$ and mass $m=3.330(1)$ g,
and an  air blower generating a nearly circular stream produced with a Pasco Variable Output Air Supply SF-9216. The mean air velocity and temperature were previously measured with a hot wire anemometer CEM DT-8880. 
The copper plate is hung vertically from a support so that the air stream is incident parallel 
to the face of the plate.
In the center of the plate a circle was painted with white-out correction fluid, that has a
thermal emissivity close to one, to ensure accurate temperature measurements with the FLIR ONE
thermal camera that was attached to a smartphone with the app Vernier Thermal Analysis Plus 
for Flir One.

The experiment begins by heating the copper plate on a hotplate until a temperature higher  than $150^o$C which is the highest value that the thermal camera can measure with the  Vernier app. Since the heat conduction inside is very fast and in less than a second it becomes uniform, it is not crucial if the temperature of the plate is not uniform during this process. 

After reaching this temperature the plate is placed on the support where the air stream hits as indicated in Fig.~\ref{fig:montaje}. We used the thermal camera to record a video with 9 frames per second and measured the plate temperature in each one while it was cooled until $25^oC$. The process was repeated for different stream velocities: $v=[(5.5\pm 0.5), (11\pm 1), (13\pm 1), (18\pm 2), (22\pm 2)]$m/s. The uncertainty is obtained as the standard error of the fluctuations of the anemometer reading.
In all the measurements considered here the stream temperature was in the range between $20^o$C and $25^o$C, so $\mathrm{Pr}\approx0.71$ and condition given by  Eq.~\ref{eqpr} remains valid.

\section{Results and discussion}
\begin{figure}
    \centering
    \includegraphics[width=.95\columnwidth]{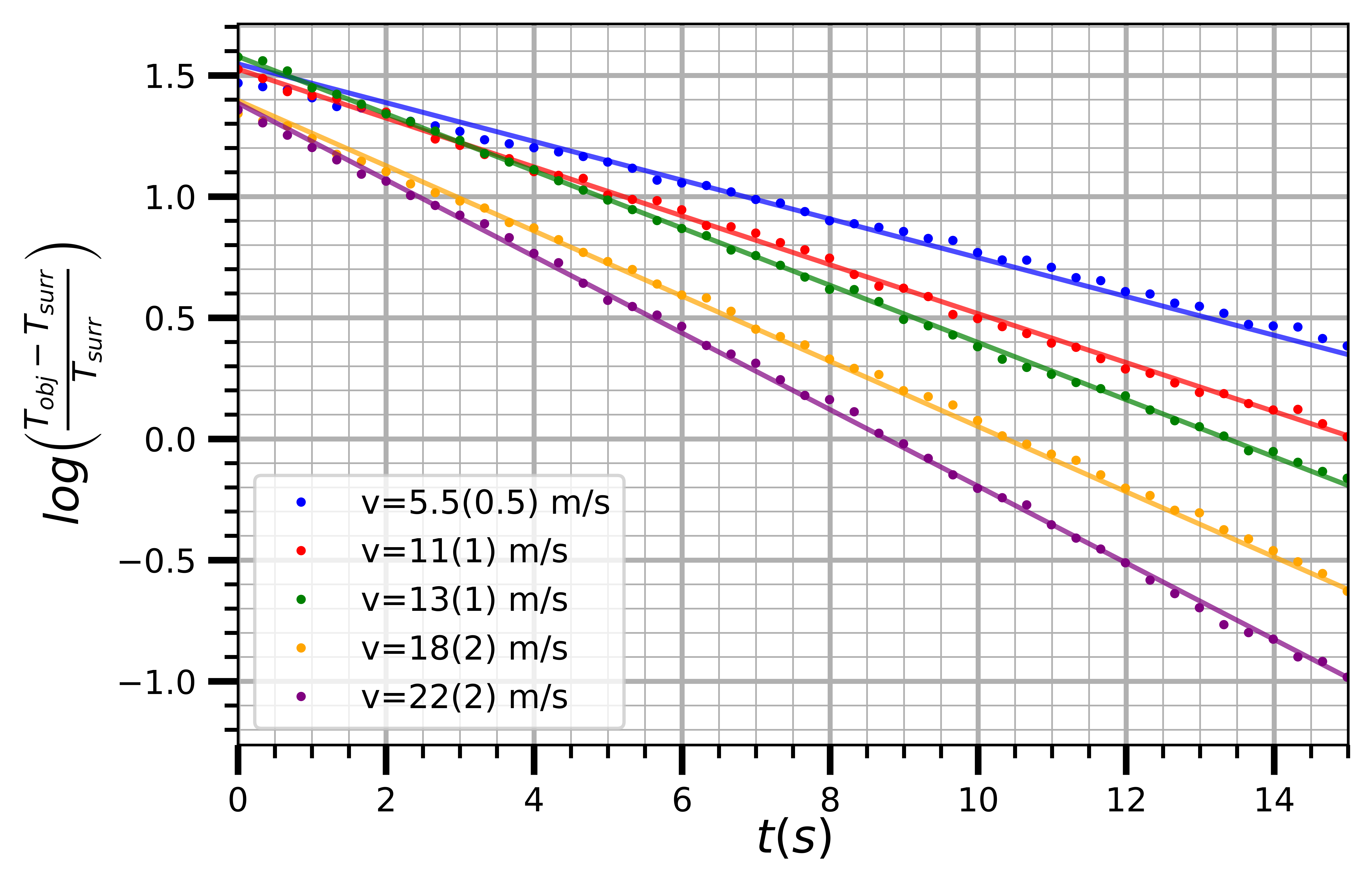}
    \caption{Cooling process of the plate. The points
    correspond to the dimnensionless temperature as a
    function of the mean stream velocity (indicated in the legend box). The lines represent linear fits of the experimental values. The correlation coefficient is $r=0.997$ for $v=5.5$m/s, and $r=0.999$ for the other mean velocity values. The linear relationship reveals that the radiation heat transfer can be neglected in this experiment.}
    \label{fig:linealizacion}
\end{figure}

Using the thermal camera record we obtained the temperature of the plate as it cooled over time for each air velocity, and calculated $\log(T_{o}-T_{e})$ in order to analyze the consistency of the experimental data with the linearization of Eq.~\ref{eqTau}. Fig.~\ref{fig:linealizacion} shows this relation, and the linear fit of the experimental values for each cooling. Given that the correlation coefficient is $r=0.997$ for $v=5.5$m/s, and $r=0.999$ for the other velocities, is not necessary to take into account the radiation heat loss. We even compared the two terms on the right hand side of the Eq.~\ref{eqRadiation} for different temperature values, and in all the cases we obtain that $\alpha_{conv}\cdot A\cdot (T_{o} - T_{e})>>\varepsilon\cdot\sigma\cdot A\cdot (T_{o}^4-T_{e}^4)$.

We obtain $\tau$ for each cooling from the slope of the linear fit in Fig.~\ref{fig:linealizacion}. Then, we calculated the $\alpha_{conv}$ from Ec.~\ref{eqTau} and the maximum value was $\alpha_{conv-max}=253$W/K. The conductive heat transfer coefficient can be calculated as $\alpha_{cond}=\lambda/s=(398$W/Km$)/(0.02$m$)=19990$W/K, where $\lambda$ is the thermal conductivity and $s$ the large of the plate. Consequently, $\mathrm{Bi}_{max}=\alpha_{conv-max}/\alpha_{cond}=0.013<<1$ so we can consider the object as homogeneously heated.

\begin{figure}
    \centering
    \includegraphics[width=.98\columnwidth]{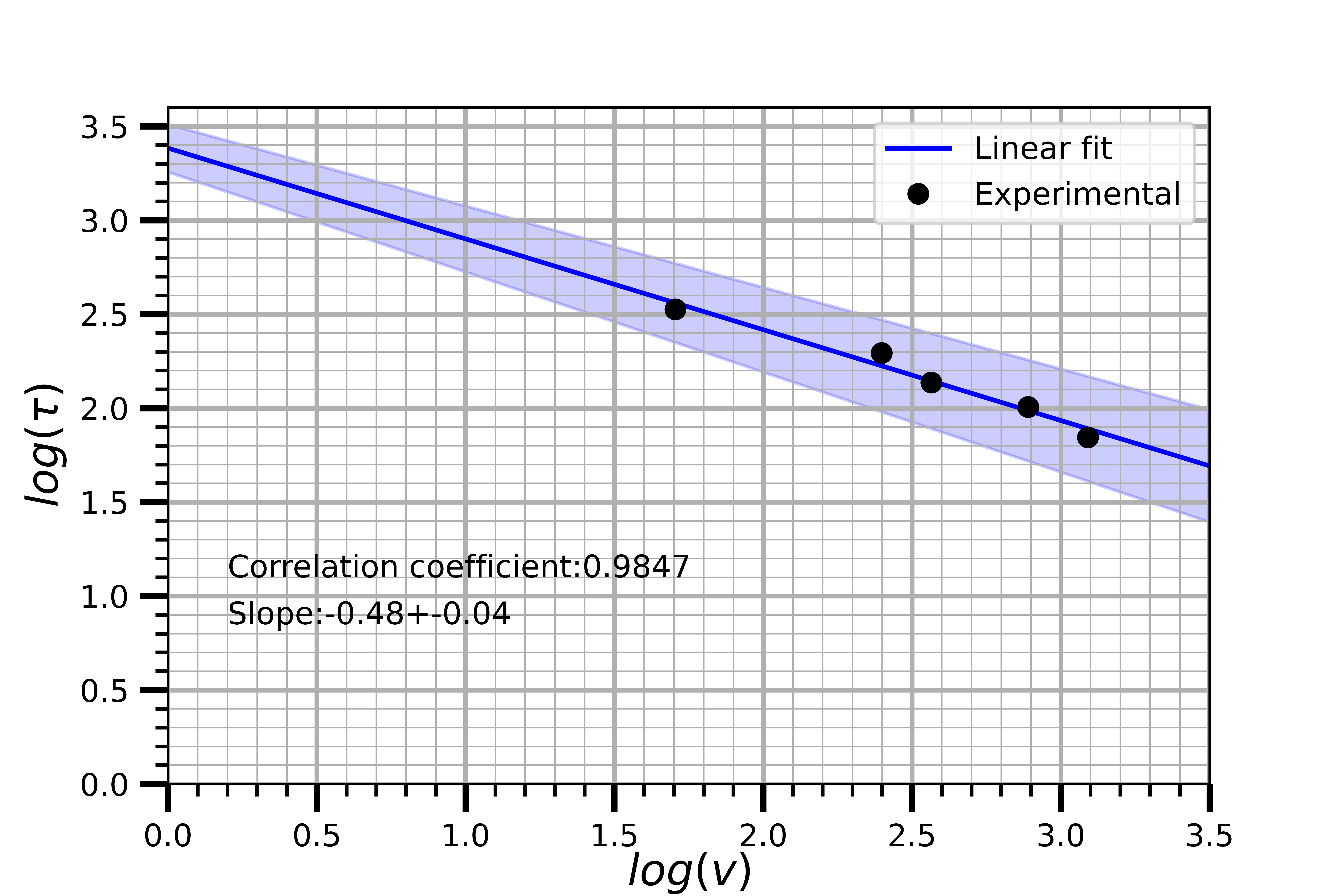}
    \caption{Log-log plot of the characteristic cooling time, $\tau$, as a function of mean stream velocity. The blue line represents the linear fit of the experimental values, obtained through the least squares method. Furthermore, the shaded blue area represents the 95\% confidence interval of this fit. The slope of this graph reveals  that  the relationship $\tau \propto v^{-1/2}$ is reasonably valid within the margins of uncertainty.
}
    \label{fig:tau}
\end{figure}

Figure~\ref{fig:tau} shows the relation between the characteristic cooling time and the averaged velocity of the air stream. The slope of the linear fit was obtained through the least squares method and is $0.48(4)$. This value is coherent with that mentioned in Eq.~\ref{eqVair}.  The uncertainty region of the linear fit is also shown in shading blue in the graph.  The correlation coefficient obtained shows that the quality of the linear fit is very acceptable.

The dependence of the characteristic cooling time on the stream velocity obtained is consistent with the conclusions that can be drawn from dimensional analysis. This problem was considered from the dimensional approach at the beginning of the 20th century by Lord Rayleigh \cite{rayleigh1915principle} and at the time prompted a wide-ranging debate  (see  \cite{bridgman1922dimensional}).  
Inspired by Rayleigh's arguments, we consider that, in a first approximation the characteristic decay time depends on five quantities: the thickness of the plate $d$, the average stream velocity $v$, the heat capacity per unit volume of the fluid $c$, and the thermal conductivity of the fluid  $\kappa$, and the temperature difference between the plate and the air stream $\Delta T$.
It should be noted that in this approach the air stream is considered to cover the entire plate and the temperature is homogeneous over the entire surface so that the most important variable for the cooling time is the thickness. Other properties of the fluid such as viscosity are not taken into account either, since at high Reynolds numbers, i.e. turbulent flows, they can be neglected.
Table \ref{tab:dim}  enumerates the six dimensional variables and the four imposed  dimensions.

\begin{table}
    \centering
    \begin{tabular}{|c|c|c|} \hline
    Dimensional variables & Symbol   & Imposed dimensions \\ \hline
        Characteristic time  & $\tau$ & $T$ \\ 
      Plate thickness   & $d$  & $L$ \\
       Velocity of the stream  & $v$ &  $LT^1$\\
     Volumetric heat capacity of the fluid    &  c &  $H L^{-3} \theta^{-1}$\\
       Thermal conductivity of the fluid  &  $\kappa $ &   $H L^{-1} T^{-1} \theta^{-1}$\\
       Temperature difference  & $\Delta T$ &  $\theta$ \\ \hline
    \end{tabular}
    \caption{The six dimensional variables and the four imposed dimensions. Note the dimensions of head and  temperature are denoted 
    as  $H$ and $\theta$ respectively.   }
    \label{tab:dim}
\end{table}

Therefore as a consequence of Buckingham's Pi theorem the problem depends only on two dimensionless numbers and the characteristic time can be expressed as \cite{lemons2017student}
\begin{equation}
\tau  \frac{\kappa}{c d^2} =   f \left(\frac{vcd}{\kappa} \right)
\end{equation}
where $f$ is an arbitrary function expressing the relationship between two dimensionless numbers.
Note that the temperature difference does not enter into this result. This fact is compatible with an exponential decay of the temperature difference.

If we consider here the experimentally obtained dependence with $\tau\propto v^{-1/2}$ we can further refine the above result by expressing the cooling time as
\begin{equation}
\tau =  \bigg(\frac{c}{\kappa}\bigg)^{1/2} d^{3/2} v^{-1/2}
\end{equation}

This result not only includes a velocity dependence compatible with the ex\-per\-i\-men\-tal results but also with the other quantities. Indeed, it is to be expected that the cooling time increases with the heat capacity of the fluid (i.e. air) and with the thickness of the plate and in turn decreases with the thermal conductivity of the fluid. This result is also consistent with the approach proposed by Boussinesq and mentioned by Rayleigh in the 1915 paper \cite{rayleigh1915principle}.

\section{Conclusions}

In this work we experimentally study the forced convection of a metal plate placed in front of an air stream. The plate was heated to a predetermined temperature and by means of a thermal camera we studied the characteristic cooling time in relation to the average velocity of the air stream.
The experimental results reveal that $\tau$ for each air velocity and the results were consistent with $\tau\propto v^{-1/2}$.

The experiment holds significant potential to include in thermodynamics laboratory courses. It is relatively cost-effective as it makes use of an affordable thermal camera, which enables its execution.
The problem addressed has also its importance from the point of view of the daily experience of students because in many real life situations we face cooling problems and in many of them convection plays an important role. 

The simplicity of the experimental device, which requires only a thermal camera (which can be a device that attaches to a smartphone) and other materials available in many laboratories, allows the experiment to be performed in various socioeconomic contexts. The only caution to be taken is the handling of hot objects, which must be done taking the necessary precautions.
We believe that this experiment can be a valuable and motivating resource.

\section*{Acknowledgment}
The authors would like to thank PEDECIBA (MEC, UdelaR, Uruguay).

\section*{References}

\providecommand{\newblock}{}

\end{document}